\documentclass[final,5p,times,twocolumn]{elsarticle}

\usepackage{graphicx}
\begin{document}

\begin{frontmatter}



\title{Impact-induced collapse of an inclined wet granular layer}


\author[label1]{Shinta Takizawa}

\author[label1,label3]{Hirofumi Niiya}

\author[label2]{Takahiro Tanabe}

\author[label2]{Hiraku Nishimori}

\author[label1]{Hiroaki Katsuragi} 

\address[label1]{Department of Earth and Environmental Sciences, Nagoya University, Nagoya, Aichi 464-8601, Japan}
\address[label2]{Department of Mathematical and Life Science, Graduate school of Science,
Hiroshima University, Higashihiroshima, Hiroshima 739-8526, Japan}
\address[label3]{Center for Transdisciplinary Research, Niigata University, Niigata 950-2181, Japan}

\begin{abstract}
The collapse of an inclined cohesive granular layer triggered  by a certain perturbation can be a model for not only landslides on Earth but also relaxations of asteroidal surface terrains. To understand such terrain dynamics,  we conduct a series of experiments of a solid-projectile impact onto an inclined wet granular layer with various water contents and inclination angles. As a result, we find two types of outcomes: ``crater formation'' and ``collapse''. The ``collapse'' phase is observed when the inclination angle is close to the maximum stable angle and the impact-induced vibration at the bottom of wet granular layer is sufficiently strong. To explain the collapse condition, we propose a simple block model considering the maximum stable angle, inclination angle, and impact-induced vibrational acceleration. Additionally, the attenuating propagation of the impact-induced vibrational acceleration is estimated on the basis of three-dimensional numerical simulations with discrete element method using dry particles. By combining wet-granular experiments and dry-granular simulations, we find that the impact-induced acceleration attenuates anisotropically in space. With a help of this attenuation form, the physical conditions to induce the collapse can be estimated using the block model.
\end{abstract}


\end{frontmatter}

\section{Introduction}
Landforms consisting of granular matter are ubiquitous in nature. Such granular terrains could be fluidized by heavy rainfalls and/or external perturbations such as earthquakes. If the perturbed granular terrains have slopes against the gravity, they could be relaxed by landslides. Snow avalanches are also known as a type of sudden collapse events which can be induced by a perturbation. Even on the surface of rocky astronomical objects, inclined terrains like crater walls can generally be found. Such inclined terrains are also subjected to the relaxation due to landslides triggered by meteor impacts. Due to the accumulation of impact-induced landslides, the shape of craters could be relaxed particularly on small asteroids~\cite{J.E. Richardson Jr}.

\par Recently, experimental studies using dry granular matter have been conducted to understand various granular phenomena~\cite{Andreotti2013}. A dry sand pile starts to flow when the surface angle exceeds a critical angle named the maximum stable angle. In this case, the fluidization is usually localized only on the vicinity of surface of sand pile~\cite{H.M. Jaeger,J.J. Roering}. When the dry granular heap is strongly vibrated, the complete fluidization resulting in nonlinear relaxation is also found recently~\cite{D. Tsuji}.
However, most of the natural terrains are more or less wet on Earth. Thus, the cohesion between particles plays an important role in landsliding dynamics. Moreover, the effect of particle-particle cohesion becomes significant for the case of small asteroids covered with particles so-called regolith. On small asteroids, the gravitational acceleration is very small. In such situation, the relative importance of cohesive effect, which originates from electrostatic and/or van der Waals forces, increases. Therefore, the experiments using only dry (non-cohesive) granular particles are insufficient to fully understand the dynamics of collapse of sloping terrains.

\par In this study, we use wet granular matter, to model the impact-induced collapse of a slope consisting of cohesive particles. The dynamics of wet granular matter is significantly different from dry one. For instance, the water content in wet granular layer causes various complex interactions between particles: the increase in cohesion due to liquid bridges and the decrease in granular friction due to lubrication effect~\cite{S. Herminghaus,N. Mitarai}. As a result, a small amount of water content strengthens the wet granular layer although the strength approaches the asymptotic value when a sufficient amount of water is added~\cite{H. Schubert,M. Scheel}. In general, strong agitation is necessary to fluidize the wet granular layer by breaking particles cohesion~\cite{M. Scheel2}. Thus, the collapse dynamics of inclined wet granular layer could be much more complex than that of dry granular layer.

\par Purpose of this study is revealing the collapse dynamics of inclined wet granular layer triggered by an external perturbation. Particularly, a solid-projectile impact is utilized for the perturbation. While this setup is a little tricky to mimic terrestrial landslide, it is relevant to impact-induced slope relaxation on small astronomical bodies and snow avalanches triggered by the fall of snow cornice. 
In addition to experiments on the vertical solid impact onto a dry granular layer~\cite{A.M. Walsh,Uehara2003,Seguin2009,H. Katsuragi} or cohesive granular layer~\cite{H. Katsuragi2}, cratering experiments of solid impact onto an inclined dry granular layer have been conducted to understand the effect of inclination angle and impact velocity on the crater shape and size~\cite{K. Hayashi,J. Aschauer}. 
Hayashi {\it et al}.~\cite{K. Hayashi} and Aschauer {\it et al}.~\cite{J. Aschauer} experimentally showed that an asymmetric crater is formed due to the collapse of crater rim following the approximately-symmetric transient crater formation. 
However, the catastrophic collapse induced by impact has not yet been investigated.
In addition, impact experiments on a horizontal wet granular layer have also been carried out to study the effect of water content and impact conditions on the crater shape and size~\cite{M. Manga,J.O. Marston,H. Takita}.
However, these experiments have not simultaneously considered the effects of both the cohesion and inclination, and they have not measured the propagation of impact-induced vibration which could be a key factor to trigger the catastrophic collapse. 
In this study, we perform the experiment of a solid-projectile impact onto an inclined wet granular layer with various water contents and inclination angles. 
In the experiment, we observe the response of wet granular layer (crater formation and/or collapse) and measure the vibration generated by impact. Moreover, we carry out numerical simulations with discrete element method (DEM) to formulate the vibration propagation in dry granular matter and compare it with wet granular experiment. The elastic-wave propagation in granular matter has also been an important topic to characterize the physics of granular matter~\cite{Liu1993,Jia1999,Jia2004,Brunet2008,Owens2011,Guldemeister2017,Bachelet2018}. Based on the experiments and simulations, we discuss the collapse condition of inclined wet granular layer due to the impact. 

\section{Experiment}
\begin{figure}[thb]
 \begin{center}
 \includegraphics[width=8cm]{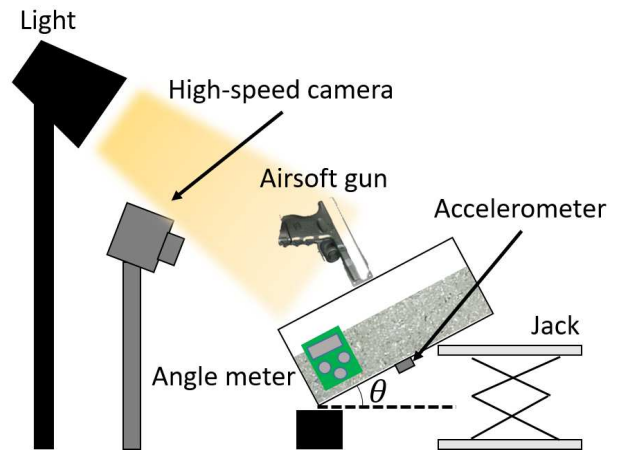}
 \caption{
Schematic of experimental setup.
The wet granular layer is poured into the container, and the surface is set to be parallel to the bottom of container.
The inclination angle $\theta$ is varied by a jack. A high-speed camera and a light are set in front of the inclined wet granular layer.
}
 \label{fig:experiment}
 \end{center}  
\end{figure}

Figure~\ref{fig:experiment} shows the schematic of experimental setup. We prepare a wet granular matter sample by mixing glass beads of 0.495 kg (AS-ONE corp. BZ04) and water. They are manually confined and shaken 100 times in a $2\times10^{-3}$ m$^3$ bottle. The diameter of glass beads is $d_{\rm g}=0.4$ mm with 25\% dispersion. The true density of glass beads is $2.5\times10^3$ ${\rm kg/m^3}$.
After preparing wet granular matter, we pour it into an acrylic container (inner width: 98 mm, length: 148 mm, height: 78 mm). On all the inside walls, the identical glass beads are glued to make frictional boundary. The wet granular layer is initially set to be horizontal. As the initial condition, the layer thickness $Z$ and packing fraction $\phi$ are fixed at $Z=28$ mm and $\phi\simeq0.49$ in almost all experiments, whereas the water content $W$ is varied in the range of  $0\leq W \leq 0.020$. Here, $W$ is defined by the ratio between water volume and total volume of wet granular layer. The thickness ($Z=28$ mm) is sufficiently deep so that the projectile never reaches the bottom of container. Although the water content $W$ decreases with time due to the water evaporation, the variation of $W$ is smaller than 1\% during the experiment. Then, the container is inclined by a jack, and the inclination angle $\theta$ is measured by an angle meter (SK Niigata seiki Bevel Box BB-180L, resolution: $0.1^\circ$) attached on the container wall. 
\par As a solid-projectile accelerator, we use a spring-driven air gun (Tokyo Marui Gindan air gun Glock 26). This gun shots a plastic spherical projectile with mass of 0.11 g and diameter of $D=6$ mm at a speed of $V=32.0 \pm 0.6$~m/s (kinetic energy of $57 \pm 2$ mJ). The error of $V$ is computed by the standard deviation of 10-time measurements of injection speed using a high-speed camera (CASIO EX-F1) at a frame rate of 1,200 frames/s. The gun muzzle is kept approximately 10 mm away from the surface of wet granular layer, and the projectile perpendicularly collides with the center of target surface. To measure the vibration induced by the impact of solid projectile, an accelerometer (EMIC 710-D) is attached on the bottom center of the container. The sampling rate of acceleration is $5\times10^4$ Samples/s. 
The entire process of impact is recorded using the high-speed camera at a frame rate of 300 frames/s with a proper illumination.

\par The experimental protocol is as following. First, we measure the maximum stable angle $\theta_{\rm m}$ as a function of water content $W$ by gradual increasing the inclination angle $\theta$. Here, $\theta_{\rm m}$ is defined as the critical inclination angle at which the wet granular layer starts to move by the slip on the bottom of container.
Once we obtain $\theta_{\rm m}(W)$, a set of systematic impact experiments with various $W$ and $\theta$ are carried out. A fresh target layer is prepared before each impact. In this study, $W$ and $\theta$ are independently varied. 

\begin{figure}[thb]
 \begin{center}
 \includegraphics[width=8cm]{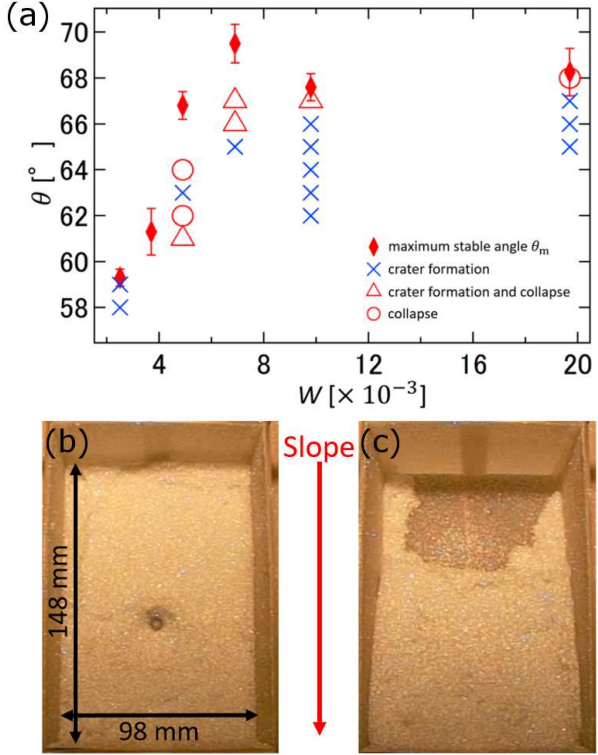}
 \caption{
(a)~Maximum stable angle $\theta_{\rm m}(W)$ (diamond symbols) and the impact outcomes (cross, triangular, and circular symbols) in $\theta$-$W$ space.
(b)~Crater formation at $W=0.0069$ and $\theta=65^\circ$. A symmetric crater is formed and left at the center of surface.
(c)~Collapse at $W=0.0069$ and $\theta=66^\circ$. The collapse completely erases the transient crater formed by the impact.
}
 \label{fig:theta_vs_W}
 \end{center}
\end{figure}

\section{Result}
Closed diamonds in Fig.~\ref{fig:theta_vs_W}(a) shows $\theta_{\rm m}(W)$ measured at $W=0.0025$, $0.0037$, $0.0049$, $0.0069$, $0.010$, and $0.020$. Each error bar indicates standard error of five measurements.
In small water content regime ($W \leq 7\times10^{-3}$), $\theta_{\rm m}$ increases with increasing $W$. However, it approaches to a constant value in larger water content regime ($W>7\times10^{-3}$). This trend of $\theta_{\rm m}$ is qualitatively consistent with previous studies~\cite{P. Tegzes,A. Samadani,S. Nowak}. $\theta_{\rm m}$ for dry ($W=0$) granular layer is approximately $25^\circ$ (not shown in Fig.~\ref{fig:theta_vs_W}(a)). In dry case, only the vicinity of surface of granular layer starts to flow. In wet situations, however, slipping of the whole granular layer on the bottom wall is observed.

\par By the impact experiments, we found two types of outcomes for the response of wet granular layer: {\it crater formation} and {\it collapse}. In this study, crater formation phase is defined by the stable crater formation without collapse (Fig.~\ref{fig:theta_vs_W}(b)), whereas collapse phase is defined by the collapse of whole wet granular layer erasing the transient crater (Fig.~\ref{fig:theta_vs_W}(c)). The crater formed in crater-formation phase is almost axisymmetric around the normal to the surface of wet granular layer. This result is contrastive to dry case in which the asymmetric crater is formed when the target surface is tilted~\cite{K. Hayashi,J. Aschauer}.
In this experiment, since the whole wet granular layer slips on the bottom of container, the container's bottom wall can be seen at the upper part in Fig.~\ref{fig:theta_vs_W}(c). 
The phase diagram of impact outcomes in $W$-$\theta$ space is shown in Fig.~\ref{fig:theta_vs_W}(a).
Cross, circular, and triangular symbols denote the crater formation, collapse, and coexistence of crater formation and collapse, respectively. This phase diagram is made on the basis of 1-3 experimental realizations in each condition.
Although there seems to be some fluctuation, the collapse phase can only be observed when $\theta$ is close to $\theta_{\rm m}$. Basically, in the small $\theta$ region, the crater formation phase can be observed. Trivially, the crater formation is always observed at $\theta=0^\circ$. This means that the increase in $\theta$ makes the wet granular layer unstable leading to collapse.

\par
\begin{figure}[thb]
 \begin{center}
 \includegraphics[width=8cm]{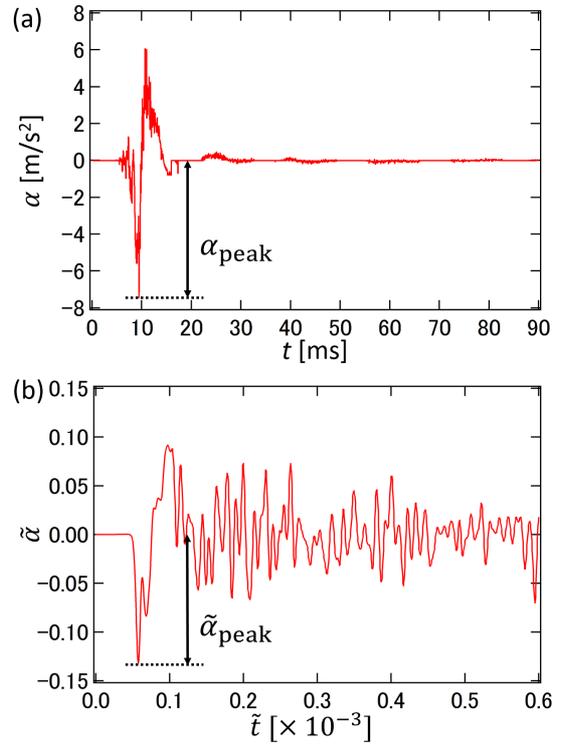}
 \caption{
(a) An experimentally measured acceleration waveform $\alpha(t)$ at the bottom center of container in the crater formation phase ($W=0.0069$ and $\theta=66^\circ$). The impact induces an impulsive signal followed by strong attenuation. (b) An example of $\tilde{\alpha}(\tilde{t},\tilde{r}=0)$ obtained by numerical simulation ($V=25$~m/s, $\tilde{z}=18$).
}
 \label{fig:acceleration_waveform}
 \end{center}  
\end{figure}

 Figure~\ref{fig:acceleration_waveform}(a) shows an example of measured acceleration $\alpha$ as a function of time $t$. The experimental conditions for the data shown in Fig.~\ref{fig:acceleration_waveform}(a) are $W=0.0069$ and $\theta=66^\circ$ corresponding to the crater formation phase.
The negative value of $\alpha$ indicates that the bottom wall experiences the acceleration towards the outside (downwards) of container. $\alpha$ is almost zero before the impact ($t<7$ ms), and it shows an impulsive signal to the negative direction by the impact (7 ms $<t<$ 10 ms). Afterwards, it exhibits a strong attenuation. To simply characterize the acceleration due to the impact, we use the peak amplitude of the acceleration $\alpha_{\rm peak}$. We use the identical projectile with identical impact velocity in all the experiments. However, the variance of measured $\alpha_{\rm peak}$ is not very small, probably due to the strong heterogeneity of wet granular layer. Specifically, the mean value of $\alpha_{\rm peak}$ for all experiments is $8.2$~m/s$^2$ and its standard deviation is $3.0$~m/s$^2$. This variation level is much greater than the level of instrumental (sensor's) uncertainty (less than a few \%). 
The noise level before impact ($t \lesssim 7$~ms in Fig.~\ref{fig:acceleration_waveform}(a)), $6$~mm/s$^2$, is three orders of magnitude less than the typical $\alpha_{\rm peak}$. Namely, the principal uncertainty in this measurement originates from the difference in structure of target granular layers among various experimental runs.

\par
\begin{figure}[thb]
 \begin{center}
 \includegraphics[width=8cm]{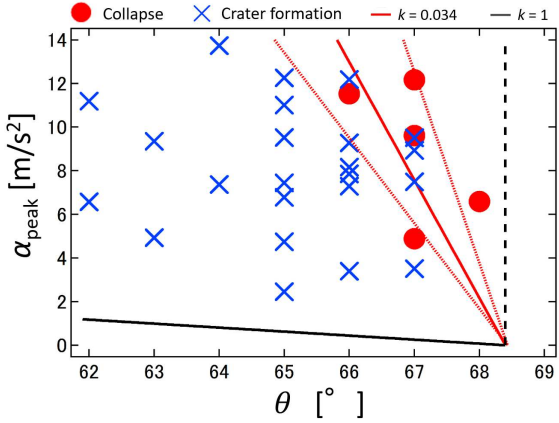}
 \caption{
Phase diagram of crater-formation phase (crosses) and collapse phase (circles) against the inclination angle $\theta$ and acceleration peak $\alpha_{\rm peak}$ acting on bottom of container.
The data in the range of $0.0069 \leq W \leq 0.020$ are used.
 Black and red lines correspond to Eq.~(\ref{eq:boundary}) with $k=1$ and $k=0.034$, respectively. The region between two dotted red lines indicates the range of $k$ estimated by uncertainties (see the main text for details).
The black dashed line indicates the average of maximum stable angle $\theta_{\rm m}=68.4^\circ$.
}
 \label{fig:apeak_vs_theta}
 \end{center}
\end{figure}

 To discuss the role of $\alpha_{\rm peak}$ for determining the impact outcomes, we make another phase diagram with $\theta$ and  $\alpha_{\rm peak}$ (Fig.~\ref{fig:apeak_vs_theta}). Here, we use the data in the range of $0.0069 \leq W \leq 0.020$, where $\theta_{\rm m}$ is almost constant value (average $\theta_{\rm m}$ is $68.4^\circ$). In Fig.~\ref{fig:apeak_vs_theta}, crosses and circles correspond to the crater-formation and collapse phases, respectively. Each plot indicates a result of one experimental realization.
These plots can roughly be separated into two groups, although they overlap at some part. In Fig.~\ref{fig:apeak_vs_theta}, we can qualitatively confirm that the large $\theta$ and/or large $\alpha_{\rm peak}$ must be fulfilled to induce the collapse.

\section{Discussion}
\subsection{Block model}
 In order to understand the criterion to induce the collapse of inclined wet granular layer due to the solid projectile impact, we propose a simple block model. In the model, the wet granular layer is assumed to be a block on the slope. Additionally, we assume that the slip condition of block corresponds to the collapse condition of wet granular layer. 
Considering the force balance among the gravity, the basal friction, and the effective acceleration driven by the impact $\alpha_{\rm eff}$, the slip condition is written as

\begin{equation}
g{\rm sin}\theta>\mu(g{\rm cos}\theta-\alpha_{\rm eff}),
\label{eq:collapse_condition}
\end{equation}
where $g$ is the gravitational acceleration, $\mu$ is the effective coefficient of friction between block and bottom wall, and $\alpha_{\rm eff}$ weakens the normal force acting on the block. 

Based on Coulomb's friction law, $\mu$ is expressed using $\theta_{\rm m}(W)$ as,
\begin{equation}
\mu={\rm tan}\theta_{\rm m}(W)
\label{eq:Coulomb's_friction_law},
\end{equation}
where $\theta_{\rm m}$ is a constant ($68.4^\circ$) since we discuss the range $0.0069 \leq W \leq 0.020$ (see diamonds in Fig.~\ref{fig:theta_vs_W}(a)).
Note that $\alpha_{\rm eff}$ should be less than $\alpha_{\rm peak}$, because $\alpha_{\rm peak}$ is measured at the center of bottom wall. Namely, $\alpha_{\rm peak}$ is the maximum value of vibrational acceleration on the bottom. However, $\alpha_{\rm eff}$ corresponds to representative average value of acceleration all over the bottom wall. The dissipation of acceleration, which is determined by characteristic features of wet granular layer, has to be properly considered to estimate $\alpha_{\rm eff}$. Here, we simply assume that $\alpha_{\rm eff}$ can be expressed using a proportional constant $k(=0\,$-1) as,
\begin{equation}
\alpha_{\rm eff}=k\alpha_{\rm peak}.
\label{eq:definition_of_k}
\end{equation}

Substituting Eqs.~(\ref{eq:Coulomb's_friction_law}) and (\ref{eq:definition_of_k}) into Eq.~(\ref{eq:collapse_condition}), the slip condition of block is rewritten as,
\begin{equation}
\alpha_{\rm peak}=\frac{g}{k}\left({\rm cos}\theta-\frac{{\rm sin}\theta}{{\rm tan}\theta_{\rm m}}\right).
\label{eq:boundary}
\end{equation}
The relationship between $\theta$ and $\alpha_{\rm peak}$ written in Eq.~(\ref{eq:boundary}) corresponds to the boundary between crater formation and collapse phases (Fig.~\ref{fig:apeak_vs_theta}).
If $\alpha_{\rm eff}$ is equivalent to $\alpha_{\rm peak}$ (i.e., $k=1$ in Eq.~(\ref{eq:boundary})), the boundary is drawn as a black line in Fig.~\ref{fig:apeak_vs_theta}. 
This line is unable to explain the experimental result. To reasonably explain the experimental result, $k$ should be much smaller than unity; $k\ll1$. The very small $k$ value reflects the strong dissipation of acceleration. 

\subsection{Estimate of model parameter $k$}
We have to understand the decay of acceleration within the impacted granular layer to quantitatively estimate the unknown parameter $k$ in the block model.
However, the decay property cannot be obtained only from the experimental data because we do not measure the acceleration inside the granular layer.
Therefore, we conduct three-dimensional (3D) numerical simulations with discrete element method (DEM). Then, the decay of acceleration due to the dissipative nature in granular matter is formulated from the numerical data. Here, we assume that the effective acceleration $\alpha_{\rm eff}$ in the block model is equivalent to the average value of $\alpha_{\rm peak}$ distribution at the bottom of granular layer, which is calculated using the specific form obtained on the basis of numerical simulation. Then, the model parameter $k$ can be estimated. In this study, we employ a simple DEM model which does not include the cohesion effect. However, as discussed later, dry (non-cohesive) DEM model is sufficient to analyze the attenuation of $\alpha_{\rm peak}$.

\par As the simulation setup, the domain shape is set to be a roofless 3D cylinder with radius of 20 $d_{\rm g}$. The granular layer is formed through the free fall of 32,768 frictionless particles with diameter of $d_{\rm g}=10$ mm and mass of 1 g.
After the free fall, the layer thickness is approximately 22 $d_{\rm g}$. Then, the identical particle ($D=10$ mm and 1 g in mass) perpendicularly collides with the surface of granular layer at two incident speeds: $V=25$, $40$~m/s. The impact point is roughly fixed at the center of surface, although it is randomly determined around the center.
We conduct 10 impact simulations at each incident speed by changing the impact point.
The detail of calculation process refers to Tanabe {\it et al}.~\cite{T. Tanabe}.

\par In these numerical simulations, we focus on the force generated by particle-particle interactions except the gravity term.
Also, we discretize radial distance from the center of cylinder $r$ and the depth $z$ ($z=0$ corresponds to the top surface of granular layer) by $d_{\rm g}$ and 2 $d_{\rm g}$ intervals, respectively. At each region, we calculate the average of vertical component of acceleration acting  on particles; and then, the maximum $\alpha_{\rm peak}(r,z)$ is computed.
To compare the experimental data with the numerical data, $r$, $z$, and $\alpha_{\rm peak}$ are normalized as,
\begin{equation}
\tilde{r}=\frac{r}{d_{\rm g}}, \;\;\; \tilde{z}=\frac{z}{d_{\rm g}},
\label{eq:dimensionless_r_and_Z}
\end{equation}
\begin{equation}
\tilde{\alpha}_{\rm peak}=\frac{\alpha_{\rm peak}}{V/T} \;\;\; (T=\sqrt{D/g}),
\label{eq:dimensionless_alpha_peak}
\end{equation}
where $T$ is the characteristic time and $D$ is the projectile diameter. An example data of $\tilde{\alpha}(\tilde{t})$ at the center ($\tilde{r}=0$) is shown in Fig.~\ref{fig:acceleration_waveform}(b), where $\tilde{t}=t/T$. Its qualitative behavior around $\alpha_{\rm peak}$ is similar to experimental result~(Fig.~\ref{fig:acceleration_waveform}(a)). Although the wave reflection and frictionless dynamics affect the relatively strong coda wave in numerical simulation, it is not important in the current study. 

\par
\begin{figure}
 \begin{center}
 \includegraphics[width=8cm]{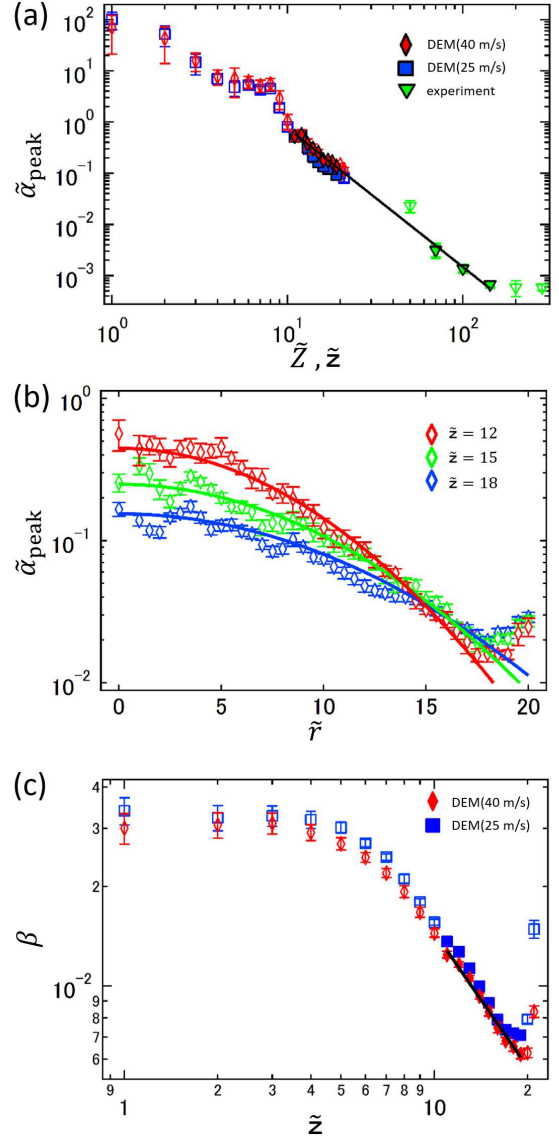}
 \caption{
(a)~Dimensionless depth ($\tilde{z}$ for simulations and $\tilde{Z}$ for experiments) dependence of the dimensionless peak acceleration $\tilde{\alpha}_{\rm peak}$. Square and diamond symbols correspond to numerical data with different incident speed $V$: $25$ and $40$~m/s, respectively. Triangular symbols are experimental data.
The black line is the fitting curve to the numerical data with $V=40$~m/s.
(b)~The radial distribution of $\tilde{\alpha}_{\rm peak}$ at $V=40$~m/s and various depths: $\tilde{z}=12$ ,$15$ and $18$ obtained by numerical simulations. Red, green and blue curves are Gaussian fits at $\tilde{z}=12$, $15$ and $18$, respectively.
(c)$~\tilde{z}$ dependency of decay parameter $\beta$ of Gaussian function. Square and diamond symbols correspond to numerical data at different incident speeds $V$: $25$ and $40$~m/s, respectively.
}
 \label{fig:DEM_graph} 
 \end{center}
\end{figure}

 First, we check the propagation of $\tilde{\alpha}_{\rm peak}$ perpendicular to the surface of granular layer at $\tilde{r}=0$ (i.e., impact direction).
To directly compare the numerical and experimental data, ${\alpha}_{\rm peak}(z)$ is also measured in the experiment with $\theta=0^{\circ}$ and $W=0.0069$ by varying the thickness of wet granular layer $Z$. We assume that the $\alpha_{\rm peak}$ measured on the bottom of thickness $\tilde{Z}=Z/d_g$ layer corresponds to the $\alpha_{\rm peak}$ at the depth $\tilde{z}=z/d_g$. 
Figure~\ref{fig:DEM_graph}(a) shows the dimensionless peak acceleration $\tilde{\alpha}_{\rm peak}$ calculated at various depths.
Square and diamond shapes denote $\tilde{\alpha}_{\rm peak}$ of simulations at incident speeds $V$ of $25$~m/s and $40$~m/s, respectively, whereas the triangle shape denotes $\alpha_{\rm peak}$ measured in the experiments (Fig.~\ref{fig:DEM_graph}(a)).
Although the focussed regions of $\tilde{z}$ (or $\tilde{Z}$) are completely separated between numerical and experimental data, they share the same power-law trend as shown in Fig.~\ref{fig:DEM_graph}(a). Note that the numerical simulation does not include the effect of wetness. Moreover, impact conditions are not identical between numerical simulations and experiments. Nevertheless, the $\tilde{\alpha}_{\rm peak}$ behavior shows a robust universal tendency. Indeed, these data are nicely fitted with the power function of $\tilde{z}$.
The obtained fitting curve is expressed as,
\begin{equation}
\tilde{\alpha}_{\rm peak}(\tilde{r}=0, \tilde{z})=\alpha_{0} \tilde{z}^{-\gamma_1},
\label{eq:alpha_peak(Z)}
\end{equation}
where $\alpha_0=4.2 \times 10^2 \pm 2.3 \times 10^2$ and $\gamma_1=2.73 \pm 0.20$ are obtained by the least square fitting.
Note that Eq.~(\ref{eq:alpha_peak(Z)}) is formulated based on the numerical data at $V=40$~m/s except $\tilde{z} \leq 10$ and $\tilde{z} \geq 20$ (open diamonds in Fig.~\ref{fig:DEM_graph}(a)), where the acceleration of particles might be affected by the free surface and the finite size of the system. 
In the experiments, $\tilde{\alpha}_{\rm peak}$ at $\tilde{Z}=50$ and $\tilde{Z} \geq 200$ also does not obey Eq.~(\ref{eq:alpha_peak(Z)}).
At $\tilde{Z}=50$ (the thinnest layer case), the solid projectile actually collides with the bottom wall. In the thick regime ($\tilde{Z} \geq 200$), $\tilde{\alpha}_{\rm peak}$ might include the effect of vibration of sidewalls since the horizontal distance from the impact point to the sidewall is shorter than the thickness of wet granular layer. It should be noticed that $\tilde{z}$ corresponds to $\tilde{Z}$ in the experimental measurements. This type of power-law decay could principally come from the geometric dissipation rather than temporal dissipation by the simple inelasticity.

Similar power-law attenuation of the impact-induced pressure was also reported in \cite{Guldemeister2017}. In general, geometric attenuation of the impact-induced pressure obeys power law. Moreover, almost the same power-law nature was obtained in both dry and water-saturated target samples~\cite{Guldemeister2017}. That is, the attenuation dynamics of impact-induced pressure (and acceleration) would be independent of wetness. More precise measurement of wave attenuation in dry and wet granular layers was also performed in \cite{Brunet2008}. They found the clear difference in dissipation between dry and wet granular layer. However, the difference is basically limited within the range of same order (small factor difference) when considering our experimental conditions (large strain by impact and low confining pressure). Such a relatively small factor difference is negligible in the logarithmic plot like Fig.~\ref{fig:DEM_graph}(a). The agreement of attenuation law between dry and wet granular matters is slightly surprising. We qualitatively consider the following reasons for this agreement. The structures of particles, by which the elastic wave propagates, in dry and wet granular layers are most likely similar. Besides, the interstitial liquid strengthens the particle cohesion but also dissipates the energy. These effects might compensate each other. As a consequence, the wave attenuation manner becomes similar between dry and wet granular matters.

 \par Next, we study the propagation of $\tilde{\alpha}_{\rm peak}$ in the radial direction using numerical results.
Figure~\ref{fig:DEM_graph}(b) shows the radial distribution of $\tilde{\alpha}_{\rm peak}$ at different depths: $\tilde{z}=12$, $15$, and $18$ with $V=40$~m/s.
These trends roughly obey Gaussian form $\sim \exp(-\beta \tilde{r}^2)$ as shown by solid curves in Fig.~\ref{fig:DEM_graph}(b) except for the outer region ($\tilde{r}>17$) where the sidewall effect is not negligible.
Figure~\ref{fig:DEM_graph}(c) shows the dependency of the fitting parameter $\beta$ on $\tilde{z}$ with different incident speeds $V$.
 Excluding the regions influenced by the free surface ($\tilde{z} \leq 10$) or bottom wall ($\tilde{z} \geq 20$) (cf. Fig.~\ref{fig:DEM_graph}(a)), $\beta$ is expressed as the power function of $\tilde{z}$ as,
\begin{equation}
 \beta = \beta_{0}\tilde{z}^{-\gamma_2},
\label{eq:beta(Z)}
\end{equation}
where $\beta_0=0.33 \pm 0.05$ and $\gamma_2= 1.36 \pm 0.05$ are obtained by the least square fitting.
Based on the above analyses, the decay of dimensionless peak acceleration driven by the impact is formulated as,
\begin{equation}
\tilde{\alpha}_{\rm peak}(\tilde{r},\tilde{z}) = \alpha_{0} \tilde{z}^{-\gamma_1}\exp\left[-\beta_{0}\tilde{z}^{-\gamma_2}\tilde{r}^2\right]\;\;\;(\tilde{z} >10).
\label{eq:alpha_peak(Z,r)}
\end{equation}
This form suggests that the attenuation of acceleration is anisotropic. We consider that this anisotropy results from the impact which has a specific direction. Some previous studies reported the exponential-type attenuation by assuming isotropic attenuation in dry granular matter~\cite{Owens2011,Bachelet2018}. In \cite{Owens2011}, however, the source of wave is not an impact, and their experimental system is two dimensional. In \cite{Bachelet2018}, while they used an impact onto a granular layer, the thickness of granular layer is much smaller than the experiment in this study. These differences are possible reasons for the variation of attenuation manners. If we restrict ourselves to the shallow region and assume the isotropic attenuation, the exponential function could also be able to explain the data behavior. However, here we employ an anisotropic attenuation because our data clearly suggest the anisotropy as shown in Fig.~\ref{fig:DEM_graph}.

Finally, we estimate the unknown parameter $k$ in the block model according to Eq.~(\ref{eq:alpha_peak(Z,r)}).
From Eq.~(\ref{eq:definition_of_k}), $k$ is defined as $k=\tilde{\alpha}_{\rm eff}/\tilde{\alpha}_{\rm peak}$. Here, $\tilde{\alpha}_{\rm peak}$ is the normalized peak acceleration at the bottom center of granular layer: $\tilde{\alpha}_{\rm peak}(\tilde{r}=0, \tilde{z})$. In addition, we assume $\tilde{\alpha}_{\rm eff}$ corresponds to the spatial average of normalized peak acceleration: $\tilde{\alpha}_{\rm eff}(\tilde{Z})=\langle\tilde{\alpha}_{\rm peak}(\tilde{r}, \tilde{Z})\rangle$.
In other words, $k$ can be estimated depending on the thickness $\tilde{Z}$ and the area of base $A$ of the granular layer.
Here, considering our experimental conditions ($\tilde{Z}=70$ and A: 98 mm $\times$ 148 mm), we obtain $k = 0.034 \pm 0.013$.
This value is able to explain the boundary between crater formation and collapse as shown in Fig.~\ref{fig:apeak_vs_theta} (red line). 
The uncertainty of $k$ value is estimated by error propagation method using the fitting uncertainties of parameters in Eq.~(\ref{eq:alpha_peak(Z,r)}) and uncertainty of $\theta_{\rm m}$ shown in Fig.~\ref{fig:theta_vs_W}(a). The vertical component of dissipation (Eq.~(\ref{eq:alpha_peak(Z)})) does not affect the estimate of $k$. Thus, the uncertainty of $k$ comes from the uncertainties of horizontal (radial) attenuation (Eq.~(\ref{eq:beta(Z)})) and $\theta_{\rm m}$. The dotted red lines in Fig.~\ref{fig:apeak_vs_theta} indicate this uncertainty.

\section{Conclusion}
We performed solid-projectile impact experiments against an inclined wet granular layer in order to understand its collapse dynamics. As a result, we found that collapse occurs when the inclination angle is close to the maximum stable angle and impact-induced peak acceleration is large enough. To explain the collapse condition, we proposed a simple block model. To quantitatively evaluate the model parameter, we numerically reproduced the vibration propagation induced by the impact by using DEM. As a result, an empirical form of peak-acceleration dissipation was obtained. Using this model, we could reproduce the experimentally obtained phase boundary between crater formation and collapse. Since the analyses were performed with dimensionless forms, it could readily be scaled up to large scale (geophysical or planetary) phenomena.

\section*{Acknowledgments}
This research has been supported by JSPS KAKENHI Nos.~15H03707 and 18H03679.





\bibliographystyle{model1a-num-names}



\end{document}